\documentclass[twocolumn,showpacs,preprintnumbers,superscriptaddress]{revtex4}
\usepackage{amsmath}
\usepackage{amssymb}
\usepackage{mathrsfs}
\usepackage{graphicx}
\usepackage{bm}
\usepackage[colorlinks=true,citecolor=blue,linkcolor=blue,urlcolor=blue]{hyperref}

\begin{document}
\title{Infrared to terahertz optical conductivity of $n$-type and $p$-type monolayer MoS$_2$ in the presence of Rashba spin-orbit coupling}

\author{Y. M. Xiao}\email{yiming.xiao@uantwerpen.be}
\address{Department of Physics and Astronomy and Yunnan Key Laboratory for Micro/Nano Materials and Technology, Yunnan University, Kunming 650091, China}
\address{Department of Physics, University of Antwerp, Groenenborgerlaan 171, B-2020 Antwerpen, Belgium}

\author{W. Xu}\email{wenxu\_issp@aliyun.com}
\address{Department of Physics and Astronomy and Yunnan Key Laboratory for Micro/Nano Materials and Technology, Yunnan University, Kunming 650091, China}
\address{Key Laboratory of Materials Physics, Institute of Solid State Physics, Chinese Academy of Sciences, Hefei 230031, China}

\author{B. Van Duppen}\email{ben.vanduppen@uantwerpen.be}
\address{Department of Physics, University of Antwerp, Groenenborgerlaan 171, B-2020 Antwerpen, Belgium}

\author{F. M. Peeters}\email{francois.peeters@uantwerpen.be}
\address{Department of Physics, University of Antwerp, Groenenborgerlaan 171, B-2020 Antwerpen, Belgium}

\date{\today}

\begin{abstract}
We investigate the effect of Rashba spin-orbit coupling (SOC) on the
optoelectronic properties of $n$- and $p$-type monolayer MoS$_2$.
The optical conductivity is calculated within the Kubo formalism. We
find that the spin-flip transitions enabled by the Rashba SOC result
in a wide absorption window in the optical spectrum. Furthermore, we
evaluate the effects of the polarization direction of the radiation,
temperature, carrier density and the strength of the Rashba
spin-orbit parameter on the optical conductivity. We find that the
position, width, and shape of the absorption peak or absorption
window can be tuned by varying these parameters. This study shows
that monolayer MoS$_2$ can be a promising tunable optical and
optoelectronic material that is active in the infrared to terahertz
spectral range.
\end{abstract}


\maketitle
\section{Introduction}
The discovery of atomically thin two-dimensional (2D) materials has
created a completely new field of research. These materials are
promising for applications in next generation of high performance
nanoelectronics devices \cite{Novoselov04}. Recently, new types of
2D materials such as monolayer transition metal dichalcogenides
MX$_2$ (M=Mo, W, Nb, Ta, Ti, and X=S, Se, Te) have been synthesized.
These 2D materials are formed by layered structures in the form of
X-M-X with the chalcogen atoms in two hexagonal planes separated by
a plane of metal atoms \cite{Wang12,Mak10,Wang14}. Transition metal
dichalcogenides (TMDCs) have a sizable band gap that can change from
indirect in multi-layers to direct in single layer structure
\cite{Wang12}. For example, molybdenum disulfide (MoS$_2$) shows a
transition from an indirect band gap of 1.29 eV in bulk to a direct
band gap near 1.90 eV in its monolayer form at the inequivalent
high-symmetry K and K$'$ points \cite{Mak10,Splendiani10}.

Since its first isolation, monolayer MoS$_2$ \cite{Mak10, Lee10} has
been investigated intensively because it exhibits interesting and
important electronic and optical properties. Very recently, one has
demonstrated that monolayer MoS$_2$ (ML-MoS$_2$) based field-effect
transistors (FETs) can have room-temperature on/off ratios of the
order of 10$^8$ and can exhibit a carrier mobility larger than 200
cm$^2$/(Vs) \cite{Radisavljevic11ACS,Radisavljevic11,
Radisavljevic13}. Other MoS$_2$ based electronic components such as
gas sensors \cite{Li12small}, phototransistors, photo-detectors with
high responsivity \cite{Yin12}, and even LEDs \cite{Sundaram13},
have been realized experimentally. Currently, the investigation of
ML-MoS$_2$ has become a fast-growing field of research with great
potential for electronics, optics and optoelectronics.

The electronic band structure for ML-MoS$_2$ can be calculated in a
$\mathbf{k}\cdotp\mathbf{p}$ theory framework \cite{Xiao12,
Kormanyos13}. Unlike massless Dirac fermions in graphene, the
electronic states in monolayer TMDCs (ML-TMDCs) can be described as
massive Dirac fermions. Furthermore, it was found that there exists
an intrinsic spin-orbit coupling (SOC) in ML-MoS$_2$, which gives
rise to a splitting of the conduction and valence bands with
opposite spin orientations \cite{Xiao12,Lu13}. In a recent work, Li
\textit{et al.} \cite{Li12} calculated the optical conductivity in
ML-MoS$_2$ in the visible range as determined by the inter-band
optical transitions from the valence to the conduction bands. The
collective excitations of ML-MoS$_2$, e.g., plasmons and screening
have also been examined and discussed \cite{Scholz13}. The presence
of a strong intrinsic SOC that couples between spin and valley
degrees of freedom has led to the proposal that the TMDCs can be
interesting materials for valleytronics and spintronics
\cite{Ganatra14, Mak12N,Zeng12N,Cao12}.

The control of carrier spin dynamics in semiconductor nanostructures
is a key issue in spintronics and it can be achieved through the
electrical manipulation of the SOC induced by the Rashba effect
\cite{Datta90,Glazov10,Wu10}. The investigation of optical
properties such as collective excitations \cite{Xu03,Wang05} and
optical conductivities \cite{Yuan05,Ang14,Azpiroz12,Azpiroz13} has
been conducted and show how the Rashba effect affects the optical
properties of traditional 2D semiconductor based spintronic devices.
This indicated that the Rashba effect has a strong influence on the
plasmon modes and low frequency optical spectrum.

When an electric field is applied perpendicular to a ML-MoS$_2$
flake, inversion symmetry is broken and, according to the Kane-Mele
model \cite{Kane05}, a Rashba SOC term is added to the Hamiltonian.
As a consequence, spin $\hat{s}_z$ is no longer a good quantum
number \cite{Kane05,Ochoa13,Klinovaja13} and it becomes possible to
have transitions between spin split electronic states. This enables
the observation of spin-related optical phenomena that were found
before to be presented in traditional 2D electron gasses (2DEGs)
\cite{Xu03,Wang05,Yuan05, Ang14,Azpiroz12,Azpiroz13}. In ML-MoS$_2$,
there is a combination of both spin-coupling mechanisms with which
one can expect to realize interesting spin-optics effects. Moreover,
thanks to the tunability of the Rashba SOC with the external
electric field, it is possible to turn these effects on and off.
Alternatively, one could also deposit various atoms on the MoS$_2$
surface that can enhance the Rashba SOC as it has been demonstrated
in gold doped graphene \cite{Marchenko12} and Bi$_2$Se$_3$(001)
surfaces \cite{King11}.

The effects of the Rashba SOC on the optical properties of graphene
have been studied \cite{Stauber09,Scholz12,Wang12jpcm,Ingenhoven10}.
It shows that the Rashba effect strongly affects the optical
properties of graphene. In the presence of the Rashba SOC, the
optical conductivity of graphene features absorption peaks and kinks
which are due to inter-band transitions between the spin split
states \cite{Wang12jpcm,Ingenhoven10}. ML-MoS$_2$ has a low energy
parabolic band which is different from the linear dispersion
relation of graphene. In order to understand the ML-MoS$_2$ material
systems more deeply and to explore their further applications in
practical devices working in the low energy bandwidth, it is
necessary to examine the roles played by the Rashba SOC in affecting
the optoelectronic properties.

Along with numerous theoretical studies, the optical and transport
properties of MoS$_2$ have also been experimentally investigated
\cite{Mak10,Splendiani10, Radisavljevic11ACS,Radisavljevic11,
Mak12N,Zeng12N,Cao12,Radisavljevic13}. However, most of these works
focus on the optoelectronic properties induced by inter-band
transitions between the conduction and valence bands that lie in the
visible range of the electro-magnetic (EM) spectrum. In this
article, we predict that the Rashba effect induces spin-flip
transitions can have a great impact on the optoelectronic response
of ML-MoS$_2$ in the infrared to terahertz range. To the best of our
knowledge, very little research has been reported on the
optoelectronic properties of ML-MoS$_2$ in this range of the light
spectrum. In this regime, the intrinsic and Rashba SOC can play an
important role in determining the optoelectronic response. We intend
studying the optical conductivity of ML-MoS$_2$ in the presence of
the Rashba SOC under linear and circular polarized radiation field.
By calculating the different contributions of intra-band and
inter-band electronic transitions, we examine the effects of $n$-
and $p$-type doping (for varying carrier density via chemical doping
or applying a gate voltage), temperature, and the Rashba SOC
strength on the optical conductivity of ML-MoS$_2$. We calculate the
electronic band structure of ML-MoS$_2$ starting from a $4\times4$
matrix Hamiltonian with the addition of the Kane-Mele Rashba SOC.

The present paper is organized as follows. In Sect. \ref{sec:band_structure},
we describe the band structure and solve for the single-particle states of ML-MoS$_2$ in the presence of the Rashba effect. The absorptive part of the optical conductivity is evaluated through the standard Kubo formalism in the presence of a linearly and/or circularly polarized radiation field in Sect. \ref{sec:optical_conductivity}. The optical transition channels for different doping types and doping levels and the results for the absorptive part of the optical conductivity are presented and discussed in Sect. \ref{sec:results}. Our main conclusions are summarized in Sect. \ref{sec:conclusion}.

\section{Electronic band structure}
\label{sec:band_structure}
In this study we consider a ML-MoS$_2$ in the $xy$-plane on top of a
dielectric wafer such as SiO$_2$ \cite{Perera13,Dolui13}. The
effective $\mathbf{k}\cdotp\mathbf{p}$ Hamiltonian for a charge
carrier (an electron or a hole) in low energy regime near the K
(K$'$)-point in ML-MoS$_2$ in the presence of the Rashba SOC can be
written as \cite{Xiao12,Lu13,Li12,Ochoa13,Klinovaja13}
\begin{align}\label{Hmt}
\hat{H}^\varsigma=&[at(\varsigma k_x\hat{\sigma}_x+k_y\hat{\sigma}_y)
+\frac{\Delta}{2}\hat{\sigma}_z]\otimes\hat{I}+\varsigma\gamma_v\frac{\hat{I}
-\hat{\sigma}_z}{2}\otimes \hat{s}_z\nonumber\\
&+\gamma_{R}(\varsigma\hat{\sigma}_x\otimes \hat{s}_y-\hat{\sigma}_y\otimes \hat{s}_x),
\end{align}
where $\hat{\sigma}_i$ and $\hat{s}_i$ are the Pauli matrices of the
sublattice pseudospin and the real spin, respectively. $\hat{I}$ is
the $2\times2$ unit matrix and the valley index $\varsigma=\pm$
refers to the K (K$'$)-valley. This Hamiltonian reads in the basis
$\Psi^{+}=\{\psi_{A\uparrow},\psi_{A\downarrow},
\psi_{B\uparrow},\psi_{B\downarrow}\}^T$ and $\Psi^{-}=\{\psi_{A\downarrow},\psi_{A\uparrow},
\psi_{B\downarrow},\psi_{B\uparrow}\}^T$ for both valleys explicitly
as
\begin{equation}\label{Hami1}
\hat{H}^\varsigma=\left(
\begin{array}{cccc}
\Delta/2 &0 &\varsigma atk_{-\varsigma} &0\\
0 &\Delta/2 &2i\gamma_R &\varsigma atk_{-\varsigma}\\
\varsigma atk_{\varsigma} &-2i\gamma_R &\gamma_v-\Delta/2 &0 \\
0 &\varsigma atk_{\varsigma} &0 &-\gamma_v-\Delta/2\\
\end{array}
\right),
\end{equation}
where $\mathbf{k}=(k_x,k_y)$ is the wave vector, $k_{\pm}=k_x\pm
ik_y$, $a=3.193$ {\AA} is the lattice parameter, $t=1.1$ eV is the
hopping parameter \cite{Xiao12}, the intrinsic SOC parameter
$2\gamma_v=150$ meV is the spin splitting at the top of the valence
band in the absence of the Rashba SOC \cite{Xiao12,Zhu11}, $\Delta =
1.66~{\rm eV}$ is the direct band gap between the valence and
conduction band used in our calculation \cite{Xiao12,Lu13,Li12}, and
$\gamma_R$ is the Rashba SOC parameter which can be tuned via an
electrical field and can be determined by \textit{ab initio }
calculation or by fitting with experimental data. Korm\'{a}nyos
\textit{et al.} \cite{Kormanyos14} estimated the value of the Rashba
parameter for ML-MoS$_2$ as $\alpha_R=0.033~$e{\AA}$^2E_z$[V/{\AA}]
for a spin split two band model which corresponds to
$\gamma_R=0.0078~$e{\AA}$E_z$[V/{\AA}] in our model with the
relation of $\gamma_R=\alpha_R\Delta/(2at)$ obtained in Ref.
\cite{Slobodeniuk16} where $E_z$ is the perpendicular electric
field. We would like to point out that the Rashba effect can not
only be tuned by a gate voltage but can also be enhanced by adatoms
as has been realized in graphene \cite{Marchenko12}. For example, a
large Rashba parameter 72 meV is found in monolayer MoTe$_2$ on a
EuO substrate \cite{Qi15}.

From the above Hamiltonian, we can describe the spin states in both
conduction and valence bands in the presence of the Rashba effect.
The corresponding Schr\"{o}dinger equation for ML-MoS$_2$ near the
valley K (K$'$) can be solved analytically and the eigenvalues are
the solutions of the diagonalized equation
\begin{equation}\label{EG}
\varepsilon^4-A_2\varepsilon^2+A_1\varepsilon+A_0=0,
\end{equation}
where
$$A_0=(\Delta^2/4+a^2t^2k^2)^2+\gamma_R^2\Delta(\Delta+2\gamma_v)-\Delta^2\gamma_v^2/4,
$$
$$
A_1=\Delta\gamma_v^2-4\gamma_v\gamma_R^2,$$ and
$$A_2=\Delta^2/2+2a^2t^2k^2+4\gamma_R^2+\gamma_v^2.$$ The energy
dispersion $\varepsilon_{\mathbf{k},\xi}=\varepsilon_{\mathbf{k},\lambda\nu}$
can be obtained through solving Eq. (\ref{EG}) analytically with the
general solution of a quartic equation \cite{Bronshtein07}. Here, we
defined the total quantum number $\xi=(\varsigma, \lambda, s)$,
where $\lambda=\pm$ refers to the conduction/valence band and
$s=\pm$ is the spin index and $\varsigma$ labels the valley. We
often use the quantity $\nu=\varsigma s=\pm$ to exploit the
valley/spin symmetry of the system.

The corresponding eigenfunction for a state near the K (K$'$) point
is
\begin{equation}
|\mathbf{k}, \xi\rangle=\psi_{\xi}(\mathbf{k},\mathbf{r})=
\mathcal{N}_{\lambda\nu}(k)[c^{\xi}_1,c^{\xi}_2,
c^{\xi}_3,c^{\xi}_4]e^{i\mathbf{k}\cdot\mathbf{r}}.
\label{eq:eigen_state}
\end{equation}
Eq. \eqref{eq:eigen_state} is expressed in the form of a row vector where the values of the eigenfunction elements are
\begin{align}
c^{\xi}_1=&-2i\gamma_Ra^2t^2k^2_{-\varsigma},
c^{\xi}_2=\varsigma atk_{-\varsigma}b^{\lambda\nu}_1,\nonumber\\
c^{\xi}_3=&-2i\varsigma\gamma_Ratk_{-\varsigma}b^{\lambda\nu}_0,
c^{\xi}_4=b^{\lambda\nu}_0b^{\lambda\nu}_2,\nonumber
\end{align}
with
\begin{align}
b^{\lambda\nu}_1=&\varepsilon_{\mathbf{k},\lambda\nu}^2-\gamma_{v}
\varepsilon_{\mathbf{k},\lambda\nu}+\Delta(2\gamma_{v}-\Delta)/4-a^2t^2k^2,\nonumber\\ b^{\lambda\nu}_2=&b^{\lambda\nu}_1-4\gamma_R^2, b^{\lambda\nu}_0=\varepsilon_{\mathbf{k},\lambda\nu}-\Delta/2.\nonumber
\end{align}
The normalization coefficient $\mathcal{N}_{\lambda\nu}(k)$ can be written as
\begin{equation}
\mathcal{N}_{\lambda\nu}(k)=1/\sqrt{h^{\lambda\nu}},
\end{equation}
with $h^{\lambda\nu}=4\gamma_R^2 a^2t^2k^2[a^2t^2k^2+(b^{\lambda\nu}_0)^2]
+a^2t^2k^2(b^{\lambda\nu}_1)^2+(b^{\lambda\nu}_0b^{\lambda\nu}_2)^2$.\par

\section{Optical conductivity}
\label{sec:optical_conductivity}
In the present study, we evaluate the optical conductivity in ML-MoS$_2$  using the standard Kubo formula\cite{Mahan90,Virosztek13}
\begin{equation}\label{Cd}
\sigma_{\alpha\beta}(\mathbf{q}, \omega)=i\frac{e^2n_i}{m\omega}
\delta_{\alpha\beta}+\frac{1}{\omega}\int^\infty_0dt
e^{i\widetilde{\omega}t}\langle[j^\dagger_\alpha(\mathbf{q},t),
j_\beta(\mathbf{q},0)]\rangle,
\end{equation}
where ($\alpha, \beta$)=$(x, y$) for a 2D system, $j_\alpha(\mathbf{q},t)$ is the current density operator, $\widetilde{\omega}=\omega+i\eta$ ($\eta\rightarrow0^+$) and $n_i$ is the carrier density for electrons in the conduction band or holes in the valence band. It should be noted that the first term in Eq. (\ref{Cd}) is the diamagnetic term \cite{Virosztek13}. In this paper we concentrate on calculating the real part of the optical conductivity, where this term does not contribute because it is purely imaginary at non-zero frequencies.

In the optical limit of $\mathbf{q}\rightarrow0$, the dynamical optical conductivity for the ML-MoS$_2$ system at an incident photon frequency $\omega$
can be written in the Kubo-Greenwood form as \cite{Virosztek13,Marder00}
\begin{align}
\sigma_{\alpha\beta}(\omega)=&
\frac{ie^2}{\omega}\sum_{\xi',\xi}\sum_{\mathbf{k'},\mathbf{k}}
\langle\mathbf{k},\xi|\hat{v}^{\varsigma}_\alpha|\mathbf{k'},\xi'
\rangle\langle\mathbf{k'},\xi'|\hat{v}^{\varsigma}_\beta|\mathbf{k},\xi\rangle\nonumber\\
&\times\frac{f(\varepsilon_{\mathbf{k},\xi})
-f(\varepsilon_{\mathbf{k'}, \xi'})}
{\varepsilon_{\mathbf{k},\xi}-\varepsilon_{\mathbf{k'},\xi'}
+\hbar(\omega+i\eta)},
\end{align}
where the velocity operator $\hat{v}^{\varsigma}_\alpha=\hbar^{-1}\partial
\hat{H}^{\varsigma}/\partial k_\alpha$, $\eta=\tau^{-1}$, $\tau$ is the transport relaxation time, and $f(\varepsilon_{\mathbf{k},\xi})=f(\varepsilon_{\mathbf{k},\lambda\nu})
=\{\exp[(\varepsilon_{\mathbf{k},\lambda\nu}-\mu_\lambda)/(k_BT)]
+1\}^{-1}$ is the Fermi-Dirac distribution function with $\mu_\lambda$ the chemical potential for electrons or holes and the temperature $T$.
It should be noted that we use a constant relaxation time for the following calculations where the specific scattering events are not considered. Very recently, the intraband optical conductivity for ML-MoS$_2$ has also been calculated \cite{Krstajic16}, and the evaluation of the relaxation time by impurity scattering had been discussed in Ref. \cite{Vargiamidis14}.

 The longitudinal optical conductivity at valley $\varsigma$ can be written as
\begin{align}\label{longitudinal}
\sigma^{\varsigma}_{xx}(\omega)=&\frac{ie^2}{\omega}\sum_{\lambda'\nu',\lambda\nu}
\sum_{\mathbf{k'},\mathbf{k}}w^{xx}_{\lambda'\nu',\lambda\nu}(\mathbf{k'},\mathbf{k})\nonumber\\
&\times\frac{f(\varepsilon_{\mathbf{k}, \lambda\nu})-f(\varepsilon_{\mathbf{k'}, \lambda'\nu'})}{\varepsilon_{\mathbf{k}, \lambda\nu}-\varepsilon_{\mathbf{k'}, \lambda'\nu'}+\hbar(\omega+i\eta)},
\end{align}
where
\begin{align}
w^{xx}_{\lambda'\nu',\lambda\nu}(\mathbf{k'},\mathbf{k})
=\frac{a^4t^4k^2[p^2+r^2+2pr\cos(2\phi)]}{\hbar^2h^{\lambda\nu} h^{\lambda'\nu'}}\delta_{\mathbf{k},\mathbf{k'}},\nonumber
\end{align}
with
\begin{align}
p=&b_0^{\lambda'\nu'}(4\gamma_R^2a^2t^2k^2+b_1^{\lambda\nu} b_2^{\lambda'\nu'}),\nonumber\\
r=&b_0^{\lambda\nu}(4\gamma_R^2a^2t^2k^2+b_1^{\lambda'\nu'}b_2^{\lambda\nu}).\nonumber
\end{align}
Moreover, the transverse or ``Hall'' optical conductivity at valley $\varsigma$ is
\begin{align}\label{Hall}
\sigma^{\varsigma}_{xy}(\omega)=&\frac{i\varsigma e^2}{\omega}\sum_{\lambda'\nu',\lambda\nu}\sum_{\mathbf{k'},\mathbf{k}}
w^{xy,\varsigma}_{\lambda'\nu',\lambda\nu}(\mathbf{k'},\mathbf{k})\nonumber\\
&\times\frac{f(\varepsilon_{\mathbf{k}, \lambda\nu})
-f(\varepsilon_{\mathbf{k'}, \lambda'\nu'})}{\varepsilon_{\mathbf{k}, \lambda\nu}
-\varepsilon_{\mathbf{k'}, \lambda'\nu'}+\hbar(\omega+i\eta)},
\end{align}
where
\begin{align}
w^{xy,\varsigma}_{\lambda'\nu',\lambda\nu}(\mathbf{k'},\mathbf{k})
&=\frac{a^4t^4k^2}{\hbar^2}
\frac{i(p^2-r^2)+2\varsigma pr\sin(2\phi)}{h^{\lambda\nu} h^{\lambda'\nu'}}\delta_{\mathbf{k},\mathbf{k'}}.\nonumber
\end{align}
The $\eta\rightarrow0^+$ limit in the above equations can be divided into a principal-value ($\mathscr{P}$) part and a Dirac-delta part through the Dirac identity
\begin{equation}
\mathrm{lim}_{\eta\rightarrow 0^+}\frac{1}{x+i\eta}
=\mathscr{P}(\frac{1}{x})-i\pi\delta(x).
\end{equation}
Thus, the optical conductivity can be separated into real and imaginary parts. The real part of the longitudinal optical conductivity and the imaginary part of the Hall optical conductivity at valley $\varsigma$ take respectively the forms
\begin{align}
\mathrm{Re}\ \sigma^{\varsigma}_{xx}(\omega)=\sum_{\lambda'\nu',\lambda\nu}\mathrm{Re}\ \sigma^{\lambda\nu\lambda'\nu'}_{xx,\varsigma}(\omega),
\label{eq:real_part_long_cond}
\end{align}
with
\begin{align}
\mathrm{Re}\ \sigma^{\lambda\nu\lambda'\nu'}_{xx,\varsigma}(\omega)=&
\int_0^\infty dk\frac{e^2 a^4t^4k^3}{2\hbar^3\omega}
\frac{(p^2+r^2)}{h^{\lambda\nu} h^{\lambda'\nu'}}\nonumber\\
&\times[f(\varepsilon_{k, \lambda\nu})
-f(\varepsilon_{k,\lambda\nu}+\hbar\omega)]\nonumber\\
&\times\delta[\omega-\omega^{\lambda'\nu'}_{\lambda\nu}(k)],\nonumber
\end{align}
and
\begin{align}
\mathrm{Im}\ \sigma^{\varsigma}_{xy}(\omega)=\sum_{\lambda'\nu',\lambda\nu} \mathrm{Im}\ \sigma^{\lambda\nu\lambda'\nu'}_{xy,\varsigma}(\omega),
\label{eq:im_part_Hall_cond}
\end{align}
with
\begin{align}
\mathrm{Im}\ \sigma^{\lambda\nu\lambda'\nu'}_{xy,\varsigma}(\omega)=&\varsigma
\int_0^\infty dk
\frac{e^2 a^4t^4k^3}{2\hbar^3\omega}\frac{(p^2-r^2)}{h^{\lambda\nu} h^{\lambda'\nu'}}\nonumber\\
&\times[f(\varepsilon_{k, \lambda\nu})
-f(\varepsilon_{k, \lambda\nu}+\hbar\omega)]\nonumber\\
&\times\delta[\omega-\omega^{\lambda'\nu'}_{\lambda\nu}(k)],\nonumber
\end{align}
where the integral of $\cos(2\phi)$ and $\sin(2\phi)$ over the angle $\phi$ are zero in Eqs. (\ref{longitudinal}) and (\ref{Hall}), respectively, and $\omega^{\lambda'\nu'}_{\lambda\nu}(k)=(\varepsilon_{k, \lambda'\nu'}-\varepsilon_{k, \lambda\nu})/\hbar$ is the energy spacing frequency between the initial and final states. \par

The total optical conductivity of the system is the summation of the contributions
at the two valleys
\begin{equation}
\sigma_{\alpha\beta}(\omega)=\sum_{\varsigma=\pm}\sigma^{\varsigma}_{\alpha\beta}(\omega).
\label{eq:sum cond}
\end{equation}
Eq. \eqref{eq:real_part_long_cond} shows that the real part of the longitudinal optical conductivity is the same for both valleys. This means that we can write the real part as $\mathrm{Re}\ \sigma_{xx}(\omega)=g_v\mathrm{Re}\ \sigma^{\varsigma}_{xx}(\omega)$ with the valley degeneracy factor $g_v=2$. For the imaginary part, Eq. \eqref{eq:im_part_Hall_cond} shows that the system is valley anti-symmetric with respect to a change in sign of $\varsigma$, i.e. $\mathrm{Im}\ \sigma^{\lambda\nu\lambda'\nu'}_{xy,+}(\omega)=-\mathrm{Im}\ \sigma^{\lambda\nu\lambda'\nu'}_{xy,-}(\omega)$. This means that the imaginary part of the Hall conductivity vanishes because the contributions from the two valleys cancel each other out.

The absorption of incoming linearly polarized radiation is given by the real part of the longitudinal conductivity. For finite frequency the optical response to circularly polarized radiation with right-handed (+) and left-handed ($-$) polarizations, the absorptive part of the optical conductivity under circular polarized radiation
at valley $\varsigma$ can be found as \cite{Li12}
\begin{equation}
\mathrm{Re}\ \sigma^{\varsigma}_\pm(\omega)=\mathrm{Re}\ \sigma^{\varsigma}_{xx}(\omega)\mp\mathrm{Im}\ \sigma^{\varsigma}_{xy}(\omega).
\label{Circular}
\end{equation}
Therefore, in order to calculate the amount of absorbed radiation, we only need to calculate the real part of the longitudinal conductivity as presented in Eq. \eqref{eq:real_part_long_cond} and the imaginary part of the Hall conductivity as given by Eq. \eqref{eq:im_part_Hall_cond}.

\section{Results and discussions}
\label{sec:results}

\begin{figure*}[t]
\includegraphics[width=14.0cm]{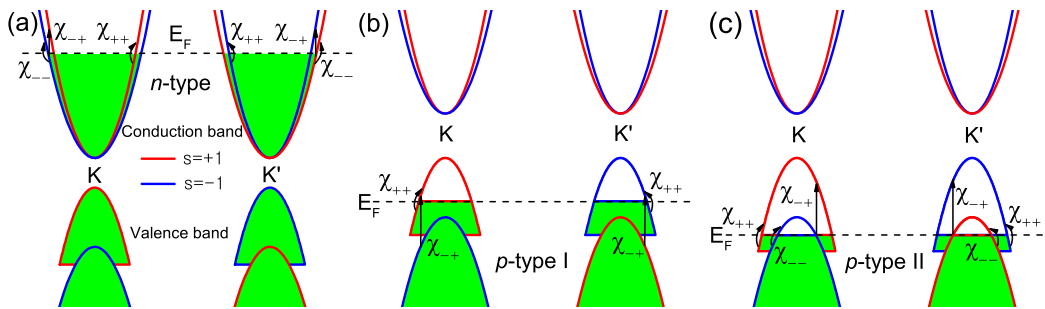}
\caption{(Color online) Schematic presentation of the band structure
and the possible optical transition channels in both valleys for
$n$-type in (a) and $p$-type I in (b) in ML-MoS$_2$ with the Fermi
level E$_\mathrm{F}$ between the two top points of spin split
valence subbands. In (c), $p$-type II ML-MoS$_2$ with Fermi energy
E$_\mathrm{F}$ for holes below the top point of $\nu=-$ valence
subband. The possible optical absorption channels are indicated by
$\chi_{\nu\nu'}$.}\label{fig1}
\end{figure*}

In this study, we consider both $n$-type and $p$-type ML-MoS$_2$ in
the presence of a relatively weak infrared or terahertz radiation
field, such that the linear response theory used in this paper is
valid \cite{Mahan90,Giuliani05}, and that the electron or hole
density do not change significantly. This is because the large band
gap suppresses the photo-excited electron-hole pairs. This means
that only the electronic transitions within the valence or
conduction band are included in the present study (i.e.
$\lambda=\lambda'$). The chemical potential $\mu_\lambda$ for
electrons in $n$-type and holes in $p$-type ML-MoS$_2$ can be
determined, respectively, through the conservation of carrier
numbers
\begin{align}\label{Ed}
n_e=g_v\sum_{\lambda=+,\nu=\pm}\sum_\mathbf{k}
f(\varepsilon_{\mathbf{k},\lambda\nu}),
\end{align}
and
\begin{align}\label{Hd}
n_h=g_v\sum_{\lambda=-,\nu=\pm}\sum_\mathbf{k}
[1-f(\varepsilon_{\mathbf{k},\lambda\nu})].
\end{align}

To perform numerical calculations for the optical conductivity, we take
the spin relaxation time for electrons $\tau^e_\mathrm{spin}$= 3 ps and for
holes $\tau^h_\mathrm{spin}$= 200 ps for spin-flip transitions \cite{Song13}.
The free carrier energy relaxation time is about $\tau_c$= 0.5 ps for
spin-conserving intra-band transitions \cite{Shi13}. Using the energy
relaxation approximation, this allows to replace the $\delta$-functions
in Eqs. \eqref{eq:real_part_long_cond} and \eqref{eq:im_part_Hall_cond} with
a Lorentzian distribution : $\delta(E)\rightarrow(E_\tau/\pi)/(E^2+E^2_\tau)$,
where $E_\tau=\hbar/\tau$ is the width of the distribution \cite{Stille12}.
It should be noted that energy relaxation time is a frequency
dependent parameter and is usually set to a constant for numerical calculation \cite{Nicol08}. The optical conductivities in Eqs. \eqref{eq:real_part_long_cond} and \eqref{eq:im_part_Hall_cond} are evaluating numerically by the standard Gauss-Kronrod quadrature method for one-dimensional integrals \cite{Piessens83} and the integrals convergence naturally due to the presence of the Fermi-Dirac function and the Lorentzian distribution of energy conservation Delta function.

In Fig. \ref{fig1}, we show the low energy electronic band structure
and corresponding optical transitions channels $\chi_{\nu\nu'}$ for
ML-MoS$_2$ at $\mathrm{K}$ and $\mathrm{K'}$ valleys. In ML-MoS$_2$,
the intrinsic SOC causes a minor spin split in the conduction band,
but it induces a large spin split in the valence band as shown in
Fig. \ref{fig1}. Furthermore, the spin-up and spin-down components
are completely decoupled, which makes electronic transitions between
opposite spin subbands impossible. Upon inclusion of the Rashba
effect, the energy spectrum is only slightly affected, but it mixes
the spin states and, therefore, the electronic transitions between
spin split subbands as shown in Fig. \ref{fig1} become possible.
Because of the absolute value of the longitudinal/Hall optical
conductivity is valley independent as we have discussed in the
previous section, we use the symbol $\chi_{\nu\nu'}$ to represent
optical transitions from $\nu$ to $\nu'$. We denote the
spin-resolved bands that are lower in energy by $\nu=-$ and the
higher energy bands by $\nu=+$.

\begin{figure*}[t]
\includegraphics[width=15.5cm]{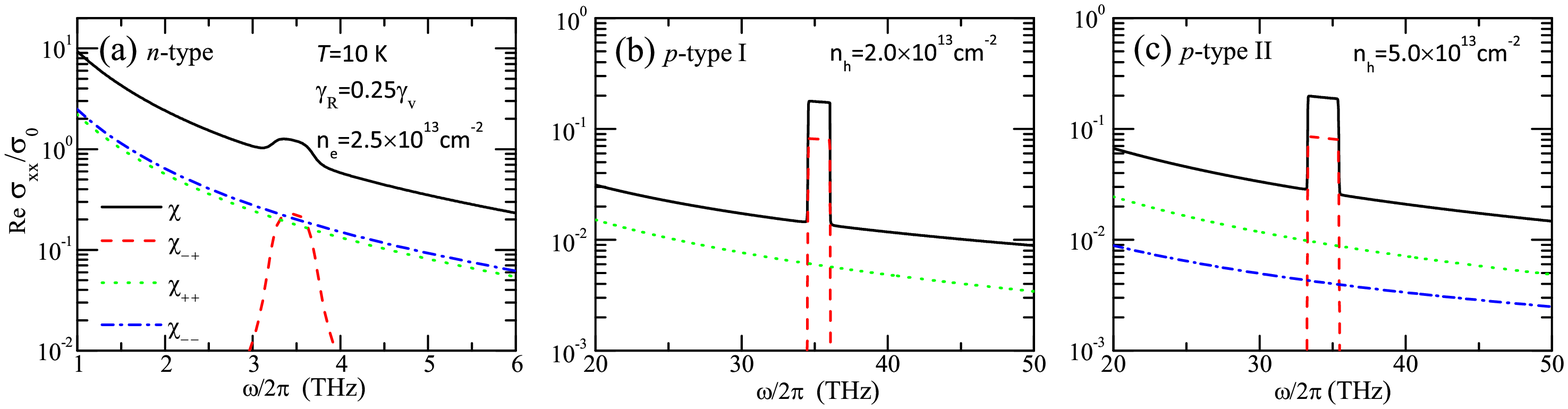}
\caption{(Color online) Finite temperature contributions from
different transition channels to the real part of the longitudinal
optical conductivity for different doping types and concentrations
as indicated. The contributions to the optical conductivity marked
in (a)-(c) correspond to the transition channels shown in Fig.
\ref{fig1} and the black solid curve presented by $\chi$ is the
summation of all the transition channels. Here,
$\sigma_0=e^2/(4\hbar)$.}\label{fig2}
\end{figure*}

ML-MoS$_2$ is a semiconductor which can be doped through techniques
such as chemical doping or applying a gate voltage
\cite{Radisavljevic13,Han12}. If the system is electron doped, i.e.,
$n$-type ML-MoS$_2$, the optical conductivity can be separated into
the contributions stemming from spin-conserving and spin-flip
transitions within the conduction band. Therefore, the conductivity
is the summation over the contributions from the optical transition
channels as shown in Fig. \ref{fig1}(a). Thanks to the large spin
split at the top of the valence band, the optical transition
channels depend strongly on the doping level in $p$-type samples.
For low $p$-type doping, the energy lower valence spin split subband
is fully occupied with electrons. Therefore, free carrier absorption
is forbidden for this subband as shown in Fig. \ref{fig1}(b). If the
$p$-type doping is larger, the subband can contribute to intra-band
transitions as indicated in Fig. \ref{fig1}(c). In order to
determine the optical absorption, one needs to calculate the real
part of the optical conductivity as outlined in the previous
section. As a consequence, measuring the optical absorption for
different doping levels allows to determine the conductivity as
defined in Eqs. \eqref{eq:sum cond}-\eqref{Circular}.

In Fig. \ref{fig2}, we show the contributions from different
electronic transition channels to optical conductivity at a fixed
electron density $n_e$ for $n$-type ML-MoS$_2$ and with fixed hole
density $n_h$ for $p$-type I/II ML-MoS$_2$ at $T= 10$ K with
$\gamma_R=0.25\gamma_v$. We see that the optical conductivity of
$n$-type ($p$-type) ML-MoS$_2$ has contributions from both
spin-conserving and spin-flip transitions as indicated by the
electronic transition channels in Fig. \ref{fig1}. Due to symmetric
dispersion relation at $\mathrm{K}$ and $\mathrm{K'}$ valleys, the
contributions to the optical conductivity in different valleys in
Fig. \ref{fig2} are identical to each other except for the fact that
the bands have opposite spin indices for the higher and lower energy
bands, respectively. In Fig. \ref{fig2}(b), free-carrier absorption
only exists in the highest spin-resolved subband because the lowest
subband is fully occupied as shown in Fig. \ref{fig1}(b). Therefore,
the intra-band transitions in $p$-type II ML-MoS$_2$ contribute more
to the optical conductivity than its $p$-type I counterpart. This
can be clearly seen in Figs. \ref{fig2}(b)-(c). We also notice that
the contributions to the optical conductivity via free carrier
absorption within $\nu=+$ subband are larger than that within
$\nu=-$ subband for $p$-type sample and it is opposite for $n$-type sample. For intra-band transitions, which mainly occurs in a small low frequency range, the relaxation time does not vary strongly with frequency. It is shown that the frequency dependence of the free carrier relaxation time in InP depends very little on frequency in low frequency range \cite{Jensen79}. At the same time, the Rashba effect will also not affect the free carrier relaxation time a lot because the Rashba effect almost does not involve the intra-band scattering mechanism. Usually, the Drude model with a single relaxation describes very well the intra-band optical conductivity \cite{Scheffler05}.

Spin-conserving intra-band transitions give rise to low frequency
THz absorption, whereas spin-flip transitions result in a wide absorption
peak as shown in Fig. \ref{fig2}(a) and a roughly rectangularly shaped
spectral absorption as presented in Figs. \ref{fig2}(b)-(c). In Fig.
\ref{fig2}(a), the total optical conductivity has a wide absorption peak
in the low frequency regime and the total optical conductivity increases
at low radiation frequency which shows the usual Drude-like behaviour.
The wide absorption peak is in the frequency regime $3-3.8$ THz which is
due to the small spin splitting in the conduction band. The absorption
peak shape is due to the smaller energy scale of the transitions in
$n$-type sample. In Fig. \ref{fig2}(b)-(c), the total optical conductivity
has an absorption window in the high-frequency regime and increases at
low radiation frequency as well. While the roughly rectangle shape absorption
windows in Figs. \ref{fig2}(b)-(c) are in the frequency regime $34.5-36$ THz
and $33.3-35.5$ THz which are induced by the larger spin splitting in the valence band.

\begin{figure}[t]
\includegraphics[width=6.5cm]{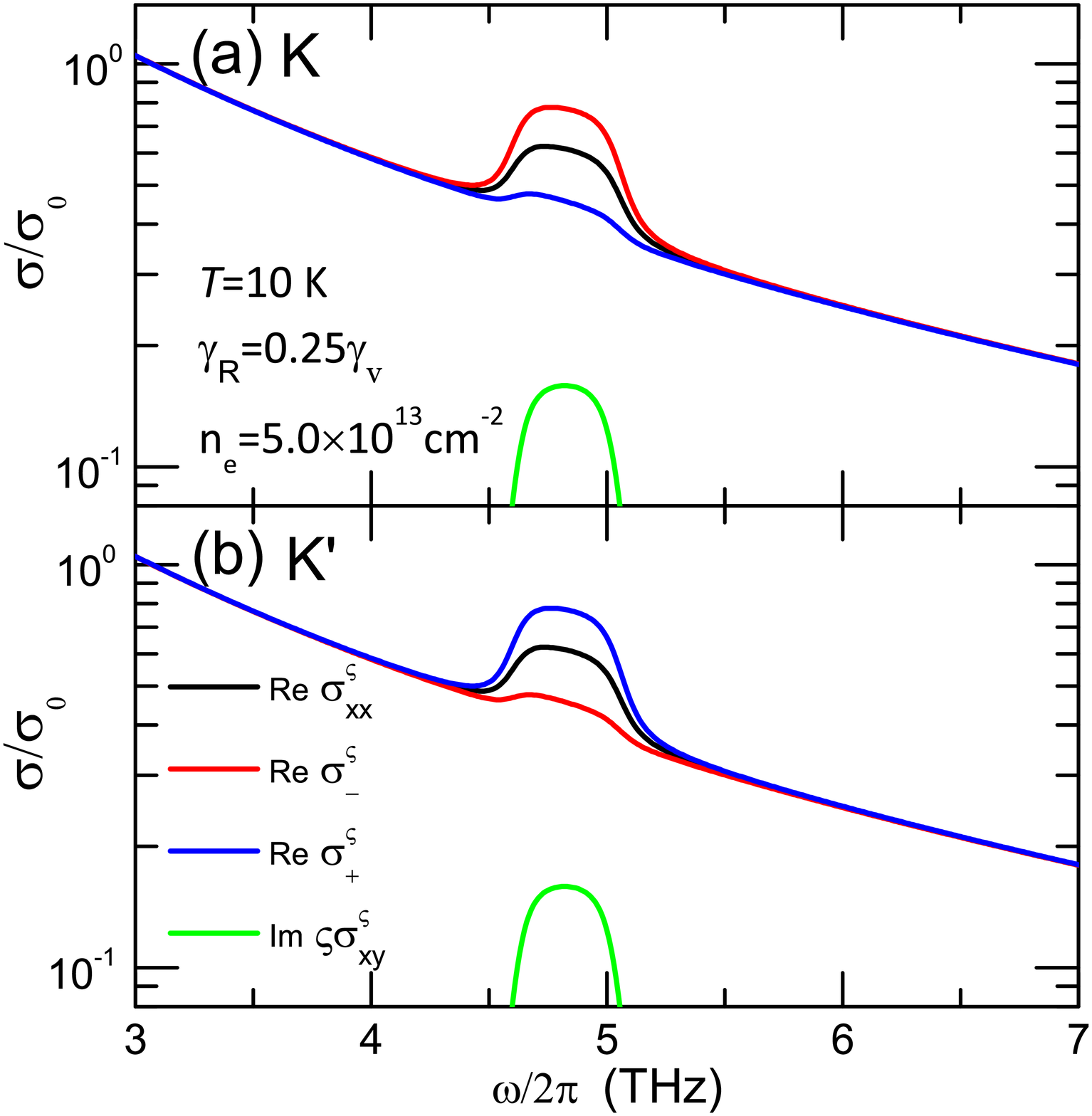}
\caption{(Color online) The absorption part of the optical
conductivity to circularly polarized radiation of $n$-type
ML-MoS$_2$ at temperature $T= 10$ K for electron density
$n_e=5\times10^{13}$ cm$^{-2}$ and Rashba parameter
$\gamma_R=0.25\gamma_v$. The results are shown for (a) the
$\mathrm{K}$ valley and (b) the $\mathrm{K'}$ valley.}\label{fig3}
\end{figure}

In Fig. \ref{fig3} we show the contributions to the optical
absorption of circularly and linearly polarized light field in the
two valleys. The optical conductivity consists of a Drude like part
induced by the contribution from intra-band transitions that
conserve spin, and a narrow absorption window due to spin-flip
transitions. Under the linearly and circularly polarized light
radiation, the Drude-like part is the same because the contributions
to the imaginary part of the Hall conductivity are zero for
intra-band transitions, which can be seen in Eq. (\ref{Hall}). As
explained in Sec. \ref{sec:optical_conductivity}, the absorption of
light with linear polarization is proportional to the longitudinal
conductivity ${\rm Re}\ \sigma^{\varsigma}_{xx}$, which is the same
for both valleys. However, the absorption of circularly polarized
light is also affected by the imaginary part of the Hall
conductivity ${\rm Im}\ \sigma^{\varsigma}_{xy}$ which is $\pm$
opposite for both valleys. As a consequence, the absorption of
circularly polarized light is valley-dependent. For the total
response to left and right handed circularly polarized light as a
function of radiation frequency, $\sigma_-$ probes mainly the
$\mathrm{K}$ valley and $\sigma_+$ the $\mathrm{K'}$ one.
Additionally, their average is equal to the longitudinal optical
conductivity. These interesting findings agree with the results for
silicene \cite{Stille12}.

\begin{figure}[t]
\includegraphics[width=6.5cm]{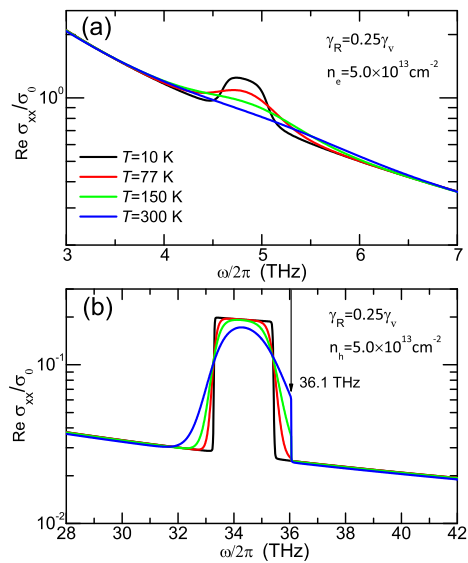}
\caption{(Color online) The real part of the longitudinal optical
conductivity as a function of radiation frequency at a fixed carrier
density (a) for $n$-type and (b) for $p$-type ML-MoS$_2$ for
different temperatures.}\label{fig4}
\end{figure}

The optical conductivity of $n$- and $p$-type II ML-MoS$_2$ are
shown in Fig. \ref{fig4} as a function of radiation frequency at
fixed carrier density and $\gamma_R$ for different temperatures. For
$\gamma_R=0.25\gamma_v$ and $T$= 10 K, the $p$-type doped sample
could be regarded as $p$-type I when the hole density
$n_h<3.36\times10^{13}$ cm$^{-2}$ and becomes $p$-type II when
$n_h>3.36\times10^{13}$ cm$^{-2}$. In Fig. \ref{fig4}(a), we find
that the strength of the wide absorption peak decreases with
increasing temperature and vanishes at room temperature. The optical
conductivity in Fig. \ref{fig4}(b) has an absorption peak at room
temperature and an absorption window at low temperature. At low
temperature, $T$=10 K, there is a sharper cutoff in absorption
window edges. We notice that a hard cutoff at 36.1 THz always exists
for the high temperature cases. With higher temperatures, the
electrons in the case of Fig. \ref{fig1}(c) redistribute their
states above the Fermi level due to the smoothing of the Fermi-Dirac
distribution. Thus, electrons can occupy some states in the top of
the $\nu=-$ subband and electronic transitions from the $\nu=-$
subband to the $\nu=+$ subband are possible. However, the top points
of the two subbands in the valence band have an energy spacing
corresponding to a radiation frequency of 36.1 THz for
$\gamma_R=0.25\gamma_v$. Thus, 36.1 THz is the largest optical
transition frequency that allows for spin-flip transitions. The
Drude-like part of the optical conductivity does not show much
differences for different temperatures because the same energy
relaxation scattering time is used in the calculation. Generally, we
can obtain a stronger absorption peak and a sharper absorption
window in the conductivity spectrum at lower temperature. However,
the spin-flip transitions induced absorption in $p$-type sample can
still be observed at room temperature. \par

\begin{figure}[t]
\includegraphics[width=6.5cm]{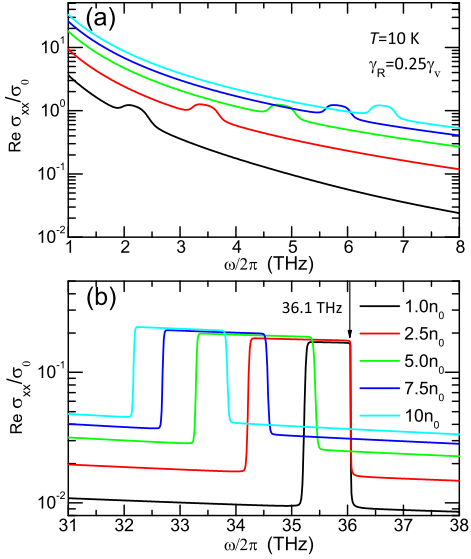}
\caption{(Color online) The real part of the longitudinal optical
conductivity as a function of radiation frequency for
$\gamma_R=0.25\gamma_v$ and $T$= 10 K and different carrier
densities where $n_0=1\times10^{13}$ cm$^{-2}$. The results are
shown for (a) $n$-type and (b) $p$-type ML-MoS$_2$.}\label{fig5}
\end{figure}
In Fig. \ref{fig5}, we show the optical conductivity of $n$- and
$p$-type ML-MoS$_2$ as a function of radiation frequency for the
fixed $\gamma_R=0.25\gamma_v$ and $T$= 10 K for different carrier
densities. For ML-MoS$_2$, $n$- and $p$-type doping samples can be
realized through the field-effect with different source and drain
contacts \cite{Radisavljevic13,Chuang14} and the doping levels can
be tuned through the application of a gate voltage. Usually, one
could reach high carrier densities \cite{Radisavljevic13,Cuong14}
and we choose the carriers densities with a magnitude of $10^{13}$
cm$^{-2}$ in our calculations. As can be seen in Fig. \ref{fig5},
the contribution to the optical conductivity from intra-band
transitions increases with increasing carrier density in both $n$-
and $p$-type samples. More interestingly, we find that the wide
absorption peak in Fig. \ref{fig5}(a) blue-shifts to higher
frequencies and the absorption window in Fig. \ref{fig5}(b)
red-shifts to lower frequencies with increasing carrier density. At
the fixed Rashba parameter and temperature, the chemical potential
for electrons/holes in $n$-/$p$-type samples increases/decreases
with increasing carrier density. As a result, the energy required
for direct spin-flip transitions increases/decreases with increasing
electron/hole density in $n$-/$p$-type ML-MoS$_2$ due to the Pauli
blockade effect \cite{Krenner06}. This is how the blue-shifts of the
absorption peak in $n$-type sample and the red-shifts of the
absorption window in $p$-type sample can occur. In Fig.
\ref{fig5}(b), we see that the right boundary of the absorption
windows at position 36.1 THz is the largest optical transition
frequency for those cases. Apart from the contribution of intra-band
transitions, we also find that the strength of the absorption peak
in $n$-type samples and the height of the absorption window in
$p$-type samples is slightly affected by the carrier density.
Besides, the width of the absorption peak in $n$-type sample varies
slightly with electron density and the width of the absorption
window in $p$-type sample varies distinctly with changing hole
density. These theoretical results show that the optical absorption
of ML-MoS$_2$ in THz and infrared regime can be effectively tuned by
varying the carrier density.\par

\begin{figure*}[t]
\includegraphics[width=12.0cm]{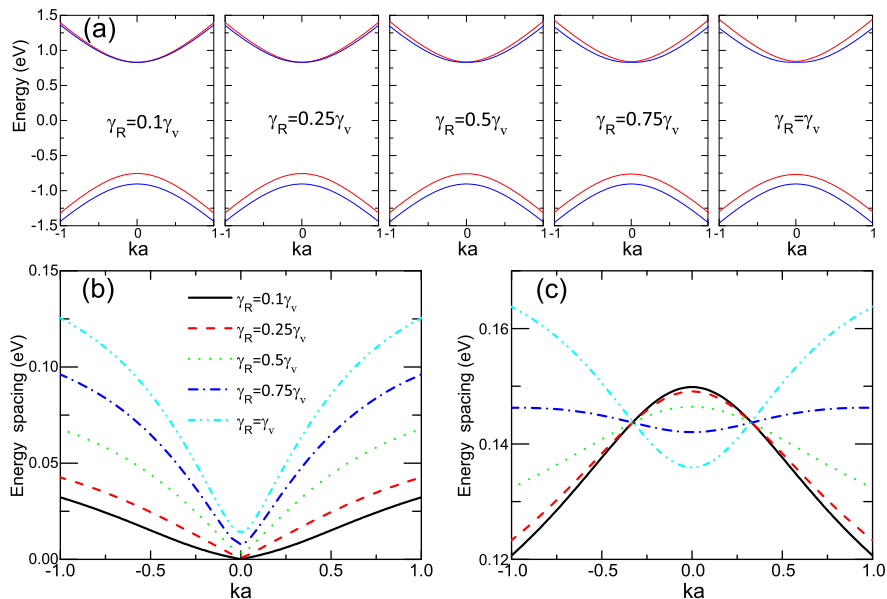}
\caption{(Color online) (a) Low-energy effective band structure of
ML-MoS$_2$ at $\mathrm{K}$ valley with different $\gamma_R$. The
spin-up ($s=1$) and spin-down ($s=-1$) subbands are denoted by red
and blue solid curves, respectively. The energy spacing between two
spin split subbands with different $\gamma_R$ are shown in for (b)
conduction band and (c) valence band.}\label{fig6}
\end{figure*}

In Fig. \ref{fig6}, we plot the low energy band structure and the energy
spacing of spin split subbands in conduction and valence bands of ML-MoS$_2$
as a function of wavevector at $\mathrm{K}$ valley in the presence of the
Rashba effect with different Rashba parameters. Usually, the Rashba SOC
strength in semiconductors can be tuned by an electric field \cite{Datta90,Glazov10,Wu10}.
In Fig. \ref{fig6}(b), near the K point, the energy spacing between two spin-orbit split conduction subbands increases with increasing wavevector or Rashba parameter. With increasing $\gamma_R$, the energy spacing turns from a roughly parabolic curve to a roughly linear line and a spin-orbit split gap can be observed at $ka=0$ for large $\gamma_R$. In Fig. \ref{fig6}(c), one can see that the energy spacing between two spin split valence subbands decreases with increasing wavevector when
$\gamma_R=0.1\gamma_v, 0.25\gamma_v$, and $0.5\gamma_v$. When $\gamma_R=0.75\gamma_v$, the energy spacing of spin split subbands first increases then decreases slightly with increasing wavevector. Whereas for $\gamma_R=\gamma_v$, the energy spacing of the subbands
increases with increasing wavevector. We find that the band structure of ML-MoS$_2$ can be fine-tuned by the Rashba effect and a more complicated band structure of spin split subbands with different Rashba parameters can be found in the valence bands.\par

\begin{figure}[b]
\includegraphics[width=6.5cm]{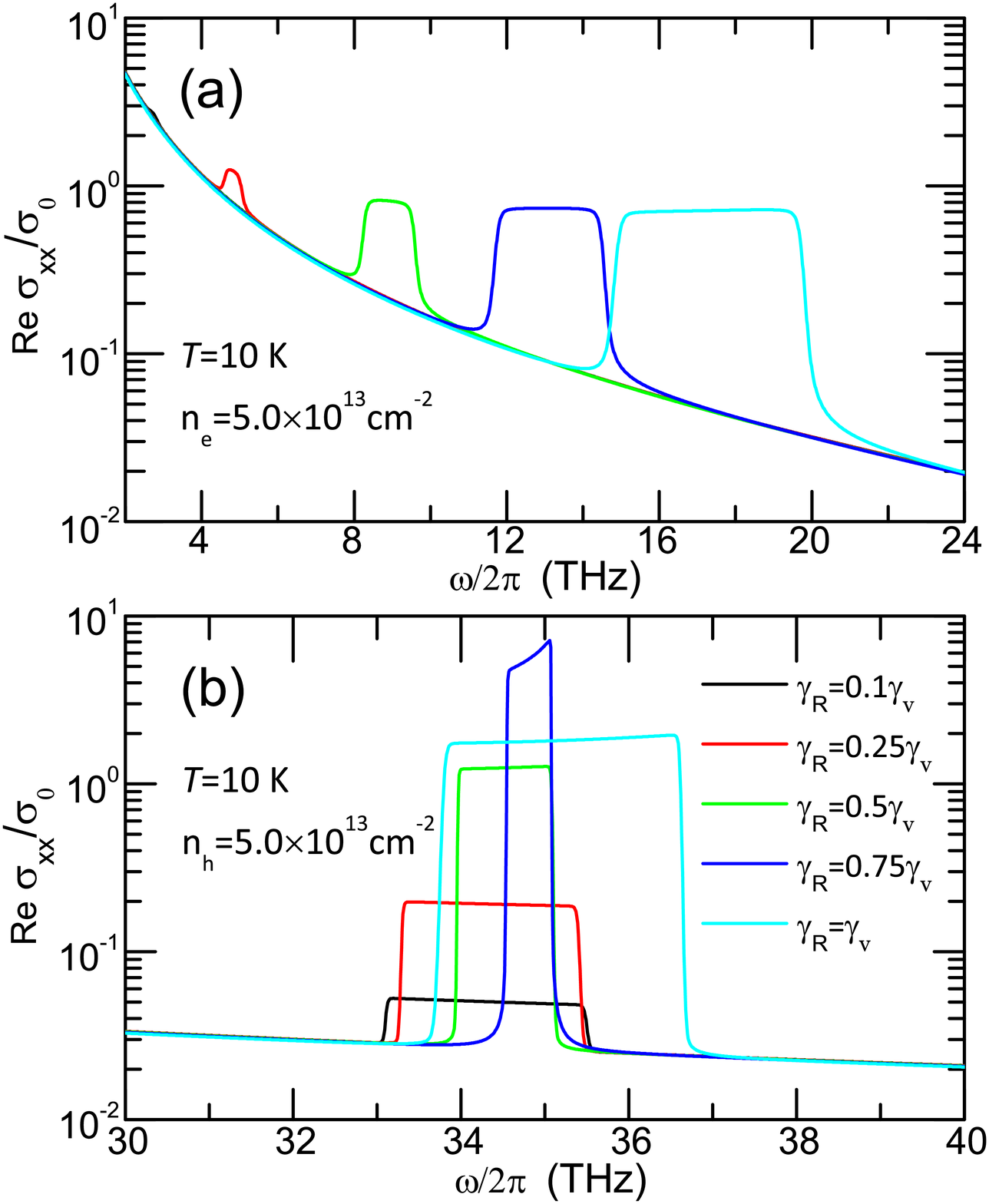}
\caption{(Color online) The real part of the optical conductivity as a function of
radiation frequency at fixed temperature $T$= 10 K and carrier density
$5\times10^{13}$ cm$^{-2}$ for different Rashba spin-orbit parameters. The results
are shown for (a) $n$-type ML-MoS$_2$ and (b) $p$-type II ML-MoS$_2$.}\label{fig7}
\end{figure}

The optical conductivity of $n$- and $p$-type ML-MoS$_2$ are shown
in Fig. \ref{fig7} as a function of radiation frequency at the fixed
temperature and carrier density for different Rashba parameters.
At large carrier density (e.g. $5\times10^{13}$ cm$^{-2}$), we
find that the top point of $\nu=-$ valence subband and the chemical
potential of electrons/holes in $n$-/$p$-type ML-MoS$_2$ do not vary
with changing $\gamma_R$. Thus, for a $p$-type ML-MoS$_2$ with a
hole density $n_h=5\times10^{13}$ cm$^{-2}$, it can be always
regarded as $p$-type II for different $\gamma_R$. With large
carrier density, we find that the dependence of the Fermi-level on the Rashba parameter is negligible. With low carrier density, we find that the Fermi level first stays the same with low Rashba coupling strength and then decreases slightly with increasing Rashba coupling strength in $n$-type sample. While in low density $p$-type sample, the Fermi level always decrease slightly with increasing $\gamma_R$.

In Figs. \ref{fig7}(a)-(b), we see that the Drude-like part of the optical
conductivity in $n$/$p$-type ML-MoS$_2$ is not affected by the value
of $\gamma_R$. The wide absorption peak in Fig. \ref{fig7}(a)
blue-shifts and the shape of it turns into an absorption window with
increasing $\gamma_R$. At the same time, the width of the absorption
window increases with increasing $\gamma_R$ and the height of the
absorption window varies slightly. In Fig. \ref{fig7}(b), we can see
that there are also absorption windows in the optical conductivity
curve with different Rashba parameters but the width and height of
them vary a lot with changing $\gamma_R$. Through the five sets of
data plotted in Fig. \ref{fig7}(b), we find that the widest
absorption window can be obtained when $\gamma_R=\gamma_v$ and the
narrowest absorption window is found when $\gamma_R=0.75\gamma_v$.
This feature can be derived from Fig. \ref{fig6}(c). We see that the
boundaries of these absorption windows in Fig. \ref{fig7}(b) do not
match the energy spacings between two spin split valence subbands at
$ka=0$ in Fig. \ref{fig6}(c). That means that the boundaries of
spin-flip transitions are mainly delimitated by the holes' optical
transitions near the Fermi level limited by the Fermi-Dirac
distribution in case of a large hole density and low temperature.
When $\gamma_R=0.75\gamma_v$, the narrow region of energy spacing in
Fig. \ref{fig6}(c) results in the narrowest absorption window in
Fig. \ref{fig7}(b) and a blade shape can be seen at the top of the
absorption window.\par

\begin{figure*}[t]
\includegraphics[width=16cm]{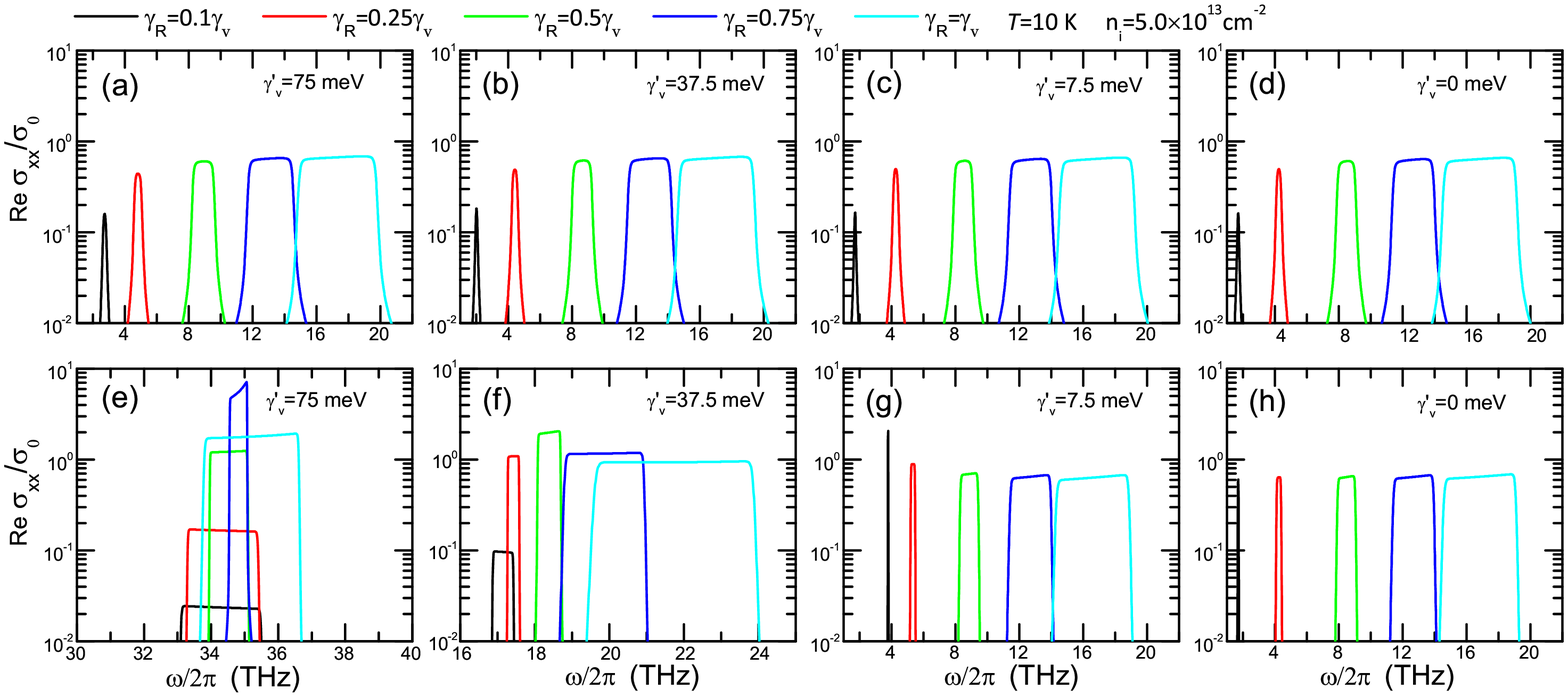}
\caption{(Color online) The real part of the longitudinal optical
conductivity contributed from the spin-flip transitions as a function of
radiation frequency at fixed temperature and carrier density for different
Rashba parameters. Here we use $\gamma'_v$ to replace $\gamma_v$ as the
intrinsic SOC parameter. The result for different $\gamma'_v$ are shown for
(a-d) $n$-type and (e-h) $p$-type samples.}\label{fig8}
\end{figure*}

Additionally, the dispersion relation for spin split conduction
subbands in ML-MoS$_2$ is similar to a traditional 2DEG system in
the presence of the Rashba effect. Thus, the results in Fig.
\ref{fig7}(a) show some similarities with traditional 2DEGs
\cite{Yuan05,Ang14}. Due to the complexity of spin-orbit split
valence subbands, the height of the absorption windows in Fig.
\ref{fig7}(b) varies a lot with changing Rashba parameter which is
different from that of traditional 2D systems. As a conclusion, we
can say that the optoelectronic properties of ML-MoS$_2$ with
different doping types can be effectively tuned by the Rashba
effect.\par

In order to understand the peculiar phenomenon that the height of
the absorption window of a $p$-type sample in the presence of the
Rashba effect varies a lot with changing Rashba parameter, we would
like to examine the role that the intrinsic SOC plays on. Although
the intrinsic SOC parameter $\gamma_v$ is a constant that cannot be
changed by external field, we would like to show how the intrinsic
SOC decides the optical property of ML-MoS$_2$ by choosing different
intrinsic SOC parameters. In Fig. \ref{fig8}, we plot the
contributions of the spin-flip transitions to the optical
conductivity of $n$- and $p$-type ML-MoS$_2$ as a function of
radiation frequency at fixed temperature and carrier density. Here,
we redefine $\gamma_v'$ as the intrinsic SOC parameter.
Fig. \ref{fig8}(a) shows the spin-flip optical conductivity part of Fig. \ref{fig7}(a). With small Rashba coupling strength, we see that the spin-flip contribution of electrons in $n$-type system decreases both in width and in height as the coupling strength decreases. As can be seen from Figs. \ref{fig8}(a)-(d), the behavior of the optical
conductivity with different $\gamma_R$ is only slightly affected by
the intrinsic SOC. The slight difference is caused by the minor
modification of the spin split in the conduction band with different
$\gamma_v'$. From Figs. \ref{fig8}(e)-(h), we find the following
features: (i) with large intrinsic SOC (eg. $\gamma_v'=$75 meV, 37.5
meV), the height of the absorption window vary strongly with the
Rashba parameter. (ii) The height of the absorption windows change
slightly and the shapes of these curves approach those for the
$n$-type case with a small $\gamma_v'$. (iii) In the absence of
intrinsic SOC ($\gamma_v'=0$ meV), the optical conductivity curves
in Fig. \ref{fig8}(h) is almost identical to that in Fig. \ref{fig8}(d).
The ultrafine difference is due to the different scattering times
for electrons and holes chosen in our calculation. Indeed,
the frequency and Rashba effect would also affect the relaxation time for
the inter-band transitions. As can be seen in Figs. \ref{fig8}(d) and (h), the relaxation times used in the energy relaxation approximation only slightly affect the two boundaries of the absorption window and does not affect much the total spin-flip optical conductivity curve. With larger relaxation scattering time which means smaller broadening of scattering states, the spin-flip optical conductivity will approach the result for the long wavelength optical limit. In this study, we are more interested in the spin-flip transition absorption part which is in the high frequency regime where the contribution of intra-band transitions is weak. Thus, we may safely assume that the energy relaxation time used in the numerical calculation is reliable for describing the optical absorption in the frequency range considered in this study.
Therefore, we can see that the intrinsic SOC plays an important role in affecting the optical absorption spectrum of $p$-type ML-MoS$_2$ for changing
Rashba SOC strength. In monolayer MX$_2$, the intrinsic SOC and the
other parameters in Eq. (\ref{Hmt}) vary with different compounds of
M and X. Thus, the infrared to THz optical absorption windows with
different bandwidths ranges and shapes can also be observed in other
ML-TMDCs materials in the presence of Rashba effect.\par

\section{Conclusions}

\label{sec:conclusion} In this study, we have investigated the
infrared to terahertz optoelectronic properties of $n$- and $p$-type
ML-MoS$_2$ in the presence of the Rashba effect. The optical
conductivity is evaluated using the standard Kubo formula. The
effects of the polarization of the radiation field, temperature,
carrier density, and Rashba parameter on the optical conductivity
have been examined. The total optical conductivity contains
contributions from different transition channels between different
spin states. We have also examined the role that the intrinsic SOC
plays in affecting the optoelectronic properties of ML-MoS$_2$ in
the presence of the Rashba effect. The main conclusions we have
obtained from this study are summarized as follows. \par

In ML-MoS$_2$, free carrier absorption exists in the entire infrared
to terahertz regime. Spin-flip transitions induce wide absorption
peaks and absorption windows which range from infrared to THz. Free
carrier absorption is weakly affected by the polarization direction
of the radiation, temperature, and Rashba parameter but depends
strongly on the carrier density. Under circularly polarized
radiation, the spin-flip transitions induce a valley selective
absorption. However, the summation over them is the same as the
longitudinal optical conductivity. A stronger absorption peak or
sharper absorption window can be observed at lower temperature. The
position and width of the absorption peak and absorption window can
be effectively tuned by carrier density and Rashba parameter. This
suggests that ML-MoS$_2$ has a wide tunable optical response in the
infrared to THz radiation regime.\par

In the presence of the Rashba effect, the features of optical
conductivity in $n$-type ML-MoS$_2$ are similar to those in 2DEGs
and the intrinsic SOC has a strong influence on the optoelectronic
property of $p$-type ML-MoS$_2$.

We have found that the optoelectronic properties of $n$- and
$p$-type ML-MoS$_2$ can be effectively tuned by the carrier density
and Rashba parameter which makes ML-MoS$_2$ a promising infrared and
THz material for optics and optoelectronics. The obtained
theoretical findings can be helpful for understanding of the optical
properties of ML-MoS$_2$. We hope the theoretical predictions in
this paper can be verified experimentally.

\section*{ACKNOWLEDGMENTS}
Y.M.X. acknowledges financial support from the China Scholarship Council (CSC). This work was also supported by the National Natural Science Foundation of China (Grant No. 11574319), Ministry of Science and Technology of China (Grant No. 2011YQ130018), Department of Science and Technology of Yunnan Province, and by the Chinese Academy of Sciences. B.V.D. is supported by a PhD fellowship from the Flemish Science Foundation.

\end{document}